# Ultrafast optical manipulation of atomic arrangements in chalcogenide alloy memory materials

Kotaro Makino,[1] Junji Tominaga,[2] and Muneaki Hase[1,3,*]

[1]*Institute of Applied Physics, University of Tsukuba, 1-1-1 Tennodai, Tsukuba 305-8573, Japan*
[2]*Functional Nano-phase-change Research Team, Nanodevice Innovation Research Center, National Institute of Advanced Industrial Science and Technology, Tsukuba Central 4, 1-1-1 Higashi, Tsukuba 305-8562, Japan*
[3]*PRESTO-JST, 4-1-8 Honcho, Kawaguchi 332-0012, Japan*
[*]*mhase@bk.tsukuba.ac.jp*

**Abstract:** A class of chalcogenide alloy materials that shows significant changes in optical properties upon an amorphous-to-crystalline phase transition has lead to development of large data capacities in modern optical data storage. Among chalcogenide phase-change materials, $Ge_2Sb_2Te_5$ (GST) is most widely used because of its reliability. We use a pair of femtosecond light pulses to demonstrate the ultrafast optical manipulation of atomic arrangements from tetrahedral (amorphous) to octahedral (crystalline) Ge-coordination in GST superlattices. Depending on the parameters of the second pump-pulse, ultrafast nonthermal phase-change occurred within only few-cycles (≈ 1 picosecond) of the coherent motion corresponding to a $GeTe_4$ local vibration. Using the ultrafast switch in chalcogenide alloy memory could lead to a major paradigm shift in memory devices beyond the current generation of silicon-based flash-memory.

## 1. Introduction

Chalcogenide alloys are semiconducting materials with periodic and random bonding networks, whose reversible phase transition enabled electrically driven non-volatile memory devices [1, 2]. Multi-component chalcogenides, such as Ge-Sb-Te and Ag-In-Sb-Te, are potentially used in optical data storage media in the forms of rewritable CDs, DVDs, and Blu-ray discs [2–4]. Especially, $Ge_2Sb_2Te_5$ (GST) has proven to be one of the highest-performance alloys among commercially available phase-change materials, whose reliability allows more than $10^5$ write-erase cycles [3–6]. The process of a rapid phase change involved in the writing and erasing of data is induced by the irradiation of focused nanosecond laser pulses, leading to transient temperature ramping [3–6]. Thus, the dynamics of the rapid phase change in the optical recording media is generally governed by a thermal (incoherent) process, limiting the speed of write-erase cycles to the MHz-GHz range [4, 5].

A question arising from the dynamics of the phase change in chalcogenide alloys is how fast the phase transformation between the amorphous and crystalline phases occurs. Motivated by understanding the mechanism of the rapid phase change process, extensive investigations on GST have been carried out using optical absorption [7], Raman scattering and x-ray absorption fine structure (XAFS) measurements [8], and the *ab initio* molecular dynamics and density functional calculations [6, 9, 10]. However, the kinetics of the rapid phase-change in GST alloys has still been actively debated, limiting the acceleration of its switching speed.

Coherent phonon spectroscopy (CPS) is a powerful tool to study the ultrafast dynamics of structural phase transitions occurring on ultrafast time scales, as has been applied to a variety of materials, such as semimetal and semiconductors [11, 12], and Mott insulators [13, 14]. Lattice dynamics in GST alloys and atomically controlled superlattice (SL) films was examined by using a femtosecond pump-probe technique and found that the appearance of the coherent vibrational modes was significantly modified upon the phase change [15, 16]. However, yet optical manipulation of atomic arrangements in GST materials has not been reported.

Here we demonstrate in GST superlattice (SL) [17–19] that the phase change from amorphous into crystalline states can be manipulated by controlling atomic motions through selectively exciting a vibrational mode that involves Ge atoms, i.e., the local $GeTe_4$ vibrational mode in the amorphous phase. Importantly, the phase change occurs only when the time separation between the two pump-pulses is simultaneously resonant to the local vibration. In addition, the transient frequency of the local vibrational mode initially softened on a subpicosecond time scale, followed by the quenching at different final states, depending on the fluence of the second pump-pulse, to which we interpret ultrafast crystallization.

## 2. Experimental details

Recently, Chong *et al.* proposed the superlattice-like phase-change random access memory (PCRAM) on the basis of considering the GST system as pseudo-binary GeTe and $Sb_2Te_3$ alloys, such that $Ge_2Sb_2Te_5 \Leftrightarrow (GeTe)_2(Sb_2Te_3)$ [17]. Both the faster switching time (< 5ns) and lower programming current have been found in the superlattice-like PCRAM. More recently, Tominaga *et al.* reported on the fabrication of a GST superlattice (SL) based upon a Ge flip-flop transition mechanism [8, 18, 19]. Using the GST-SL films artificially designed from GeTe sub-layers and $Sb_2Te_3$ sub-layers, they experimentally confirmed a very low (only ≈ 10 % compared to GST alloy films) power operation of the phase switching (amorphous ↔ cubic) in the GST-SL, implying that GST-SL have an ideal structure to realize the flip-flop transition of Ge atoms.

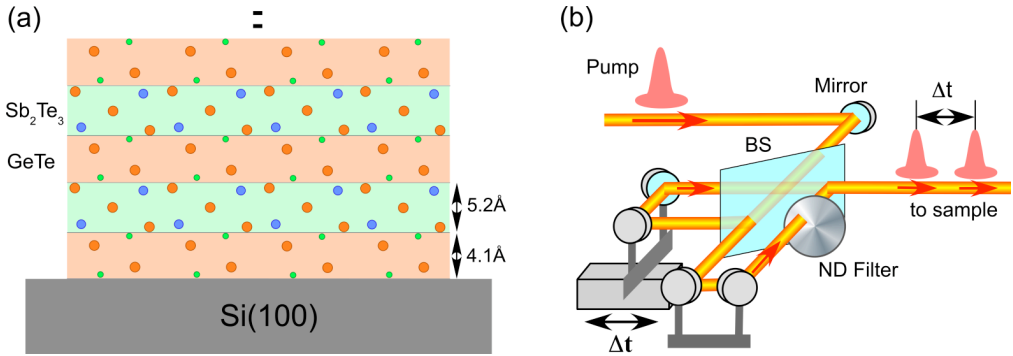

Fig. 1. (a) Schematic structure of the $Ge_2Sb_2Te_5$ superlattice, $[(GeTe)_2/Sb_2Te_3]_{n=20}$, fabricated on Si (100) substrate. (b) Optical layout of the Michelson interferometer with a piezo stage to generate the pump-pulse pair. The time interval ($\Delta t$) of the two pump-pulses is controlled by moving one arm of the interferometer. BS is a beam splitter.

A sample used in this paper was thin films of GST-SL, which consisted of twenty-repetitive sheet blocks from alternatively deposited 0.5-nm-thick GeTe and $Sb_2Te_3$ sub-layers on a Si-(100) wafer at room temperature using a helicon-wave RF magnetron-sputtering machine [Fig. 1(a)]. The phase of the as-grown GST-SL films was confirmed as amorphous by both transmission electron microscopy (TEM) and electron-diffraction measurements. By annealing the as-grown GST-SL films at 503 K for ten minutes, the GST-SL film was transformed from the amorphous into crystalline states [16, 18], where the layered structure of GST-SL films has been confirmed by high-resolution TEM measurements [19].

We utilized a 20-fs near-infrared optical pulse (1.46 eV) to excite coherent lattice vibrations in GST-SL films after injection of photo-carriers across the indirect band gap of 0.6 - 0.7 eV [7]. The photo generated carrier density was estimated to be $n_{exc} \approx 4.0 \times 10^{19}$ cm$^{-3}$, induced by the single pump-pulse with 64 µJ/cm$^2$. A couple of the pump-pulses (pump-pulse pair) were generated through a Michelson-type interferometer, where a piezo-stage was equipped under the mirrors to adjust the time interval ($\Delta t$) of the temporally separated pump-pulses to a precision of < 0.1 fs [Fig. 1(b)]. The photoinduced reflectivity change ($\Delta R/R$) was recorded as a function of the time delays between the pump and probe pulses at room temperature (295 K). We utilized a discrete wavelet transform (DWT) method [20] with a Gaussian time window to explore the transient frequency of the coherent vibrations.

## 3. Coherent phonon spectra observed in the amorphous and crystalline phases

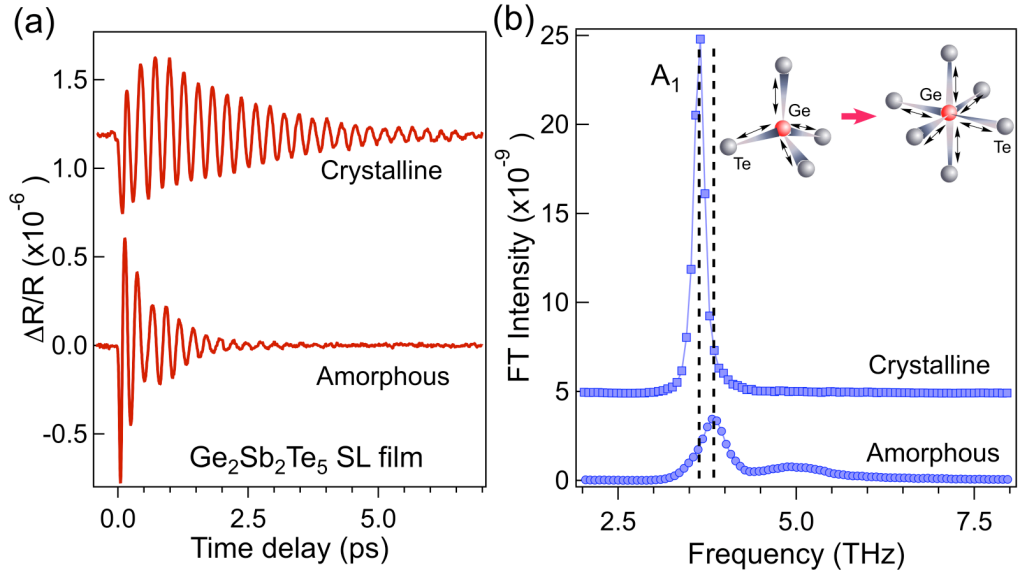

Fig. 2. (a) Transient reflectivity changes and (b) their FT spectra observed in the GST superlattice films. The dotted lines in the FT spectra represent the peak positions of $A_1$ mode. The inset of (b) represents the local structural change from GeTe$_4$ to GeTe$_6$ units.

The coherent vibration observed in the amorphous GST-SL film is compared with that observed in the crystalline phase under the single pump-pulse excitation with 64 µJ/cm$^2$ as shown in Fig. 2(a), which is the photoinduced reflectivity change ($\Delta R/R$). As seen in the corresponding Fourier transformed (FT) spectra presented in Fig. 2(b), the coherent phonon spectra obtained from the GST-SL film in the amorphous phase exhibits two peaks, one at ≈ 3.84 THz that originated from the local $A_1$ mode of the tetrahedral GeTe$_4$ species and another mode at ≈ 5.0 THz, which likely corresponds to the $A_1$ optical mode due to pyramidal SbTe$_3$ in the Sb$_2$Te$_3$ sub-layers [15, 16, 21]. Note that the stronger peak at 3.84 THz dominates the coherent phonon spectrum in the amorphous phase, unlike in Raman measurements [21] and density functional study [22], in which a broad Raman peak appeared at ≈ 4.5 THz (= 150 cm$^{-1}$). The difference in the phonon spectra would be attributed to the sample structure and the excitation mechanism of the coherent $A_1$ mode of the tetrahedral $GeTe_4$ in GST-SL; we found a pump-polarization dependence of the amplitude of the coherent $A_1$ mode (3.84 THz) (not shown). This may suggest a possibility of anisotropic excitation of carriers from bonding into anti-bonding states, which exerts a force on the lattice driving the coherent $A_1$ mode. In the crystalline phase of the GST-SL film, on the other hands, a single sharp peak appears at ≈ 3.68 THz. For $Ge_2Sb_2Te_5$ superlattice, the frequency of the local $A_1$ mode in the crystalline phase (3.68 THz) is significantly lower than that in the amorphous phase, owing to the local structural change from tetrahedral $GeTe_4$ into octahedral $GeTe_6$ species [see the inset of Fig. 2(b)]. Thus, the difference between the amorphous and crystalline phases in the GST-SL film has been clearly revealed as the peak positions of the two characteristics FT spectra.

## 4. Fluence dependence of the coherent A$_1$ modes with single pump-pulse irradiation

Before performing the present experiments using the pump-pulse pair, we first examined the pump fluence dependence of the coherent local $A_1$ mode by the irradiation of the single pump-pulse to check if the phase change from the amorphous into crystalline states is possible. The coherent phonon signal are recorded as a function of the time delay as shown in Fig. 3(a), and the corresponding FT spectra are obtained as shown in Fig. 3(b). At the lowest fluence (47 µJ/cm$^2$) the coherent phonon spectrum exhibits similar structure as in Fig. 2(b) (amorphous), where two peaks appear, one is at 3.84 THz and the other is at ≈ 5.0 THz. As the pump fluence increases, the peak frequency of the $A_1$ mode of $GeTe_4$ very slightly red-shifts from 3.84 THz to 3.80 THz, while that of the $SbTe_3$ pyramids red-shifts from ≈ 5.0 to 4.73 THz.

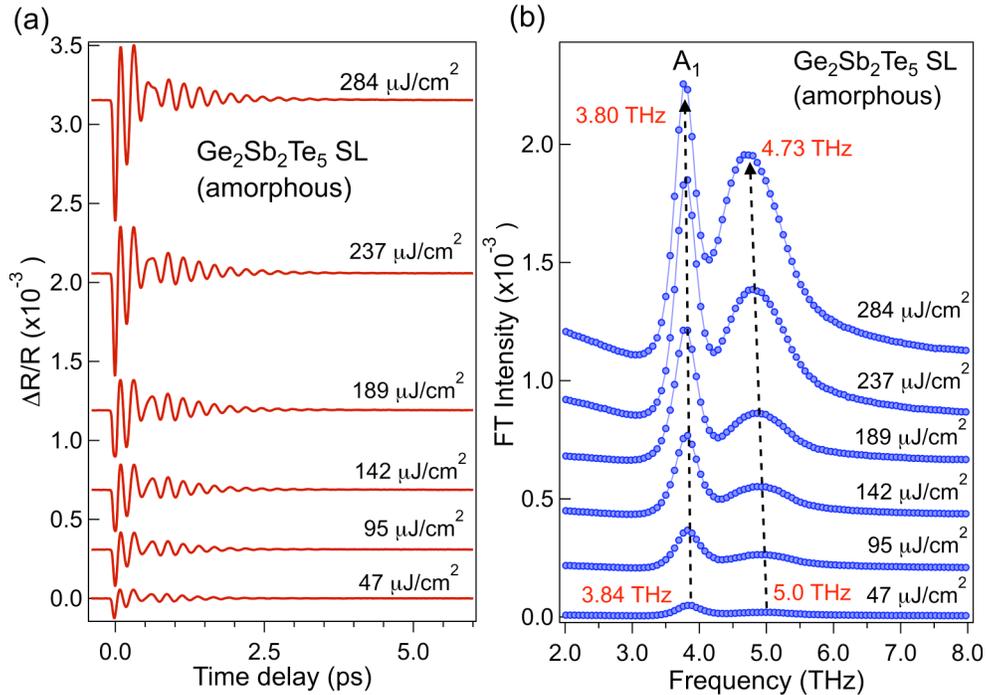

Fig. 3. (a) The transient reflectivity changes and (b) the corresponding FT spectra observed in the amorphous GST-SL at room temperature. The pump fluence was varied from 47 to 284 µJ/cm$^2$. The dashed arrows in (b) indicate the shift of the peak frequency of the $A_1$ mode, corresponding to the $GeTe_4$ local mode, and that of the pyramidal $SbTe_3$, respectively.

Moreover, it is found that the ratio of the peak intensity $I_{SbTe_3}/I_{GeTe_4}$ significantly increases with increasing the pump fluence, implying the energy deposited by the laser irradiation goes mainly into the $A_1$ mode of SbTe$_3$ pyramids. The small peak shift (only ≈ 0.04 THz) observed for the $A_1$ mode of GeTe$_4$ units cannot be accounted for the phase change since the peak frequency of the $A_1$ mode in the crystalline phase should be 3.68 THz as shown in Fig. 2(b). From the results of the single pump-pulse irradiation, we have concluded that the phase change from the amorphous into crystalline states is difficult by this single pump-pulse excitation. Thus, the experiments using the pump-pulse pair has been motivated by concentrating the energy deposited by the laser irradiation on the coherent $A_1$ mode of GeTe$_4$ in order to induce the phase change.

## 5. Coherent manipulation of coherent phonons using the pump-pulse pair excitation

*5.1 The effect of the pump-pulse pair excitation on the relaxation time of coherent phonons*

To see the effect of the pump-pulse pair excitation, we first compare the relaxation time ($\tau$) of the $A_1$ mode (due to the local GeTe$_4$ structure) observed by the single pump-pulse excitation to that by the pump-pulse pair excitation with the separation time of $\Delta t$ = 276 fs. Here, $\Delta t$ = 276 fs is close to the time period of the local $A_1$ mode of the octahedral GeTe$_6$, as is described in the next section.

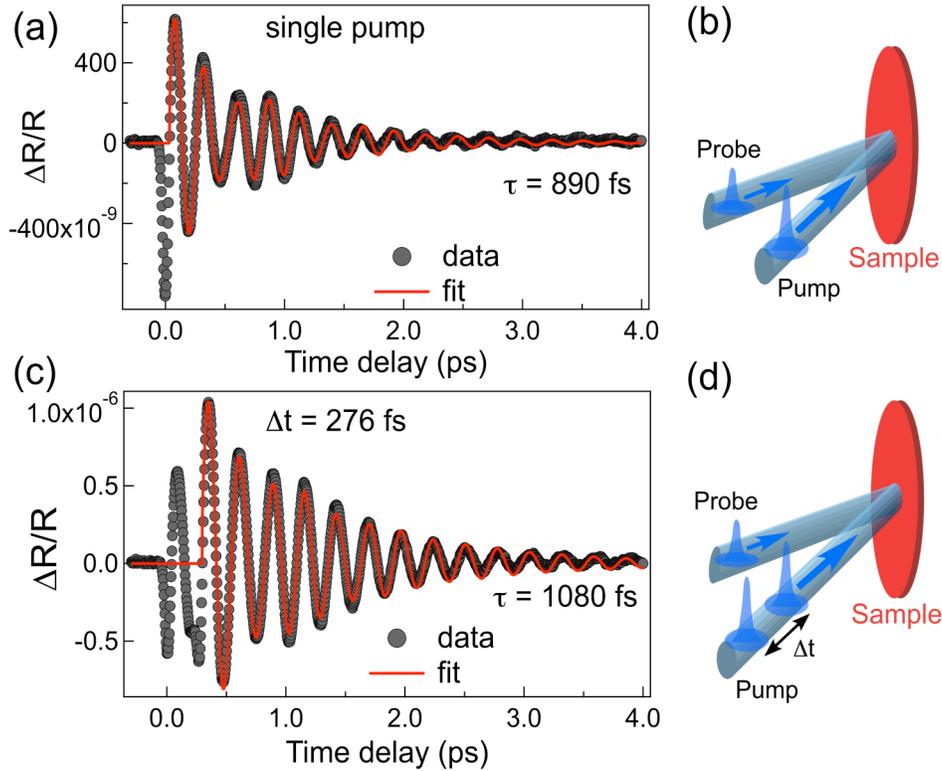

Fig. 4. (a) Transient reflectivity trace for the amorphous GST-SL film, recorded by the use of the single pump-pulse at 64 µJ/cm$^2$. (b) The experimental schemes around the sample in the case of the single pump. (c) Transient reflectivity trace for the amorphous GST-SL film by the use of the pump-pulse pair of $\Delta t$ = 276 fs, with the primary pump-pulse fluence $F_1$ = 76 µJ/cm$^2$ and the secondary pump-pulse fluence $F_2$ = 64 µJ/cm$^2$. (d) Same as (b) in the case of the pump-pulse pair excitation.

Figure 4(a) shows transient reflectivity ($\Delta R/R$) observed in the amorphous GST-SL film by the single pulse excitation [Fig. 4(b)], in which the fit of the data with damped harmonic oscillations [23] gives $\tau$ = 890 fs, while that observed by the pump-pulse pair with $\Delta t$ = 276 fs is significantly longer ($\tau$ = 1080 fs) than that in the case of the single pump-pulse [Figs. 4(c) and (d)]. Since the relaxation process of the $A_1$ mode (GeTe$_4$) in the amorphous phase has found to not depend on the lattice temperature, but depended mainly on the defect density [16], the prolongation of the relaxation time upon the pump-pulse pair excitation cannot be explained by the effect of lattice heating. Hence, this result indeed suggests a possibility of phase change from amorphous into crystalline state induced by coherent excitation because relaxation of phonon modes in crystalline phase becomes longer than that in the amorphous phase, due to the reduction of phonon-defect scattering in the crystalline phase [16, 24].

*5.2 Coherent manipulation of coherent phonons by tuning the pump fluence and time interval*

To check if the coherent excitation of the local lattice vibration induced subsequent phase change, following two procedures were examined using the amorphous GST-SL film. The first parameter varied was the fluence of the second pump-pulse ($F_2$) to determine whether critical laser fluence existed above which the phase change was induced [Figs. 5(a) to 5(c)]. In this experiment, the time interval ($\Delta t$) between the first and the second pump-pulse was fixed at 276 fs (corresponds to 3.62 THz), which is intentionally close to the time period of the local $A_1$ mode of the octahedral GeTe$_6$ (in crystalline phase), allowing the second pump-pulse to constructively enhance the vibrational amplitude toward the crystalline phase [Fig. 5(a)] [23, 25]. Most remarkably, as a result, the peak frequency significantly shifts from 3.83 THz (at $F_2$ = 0; single pump) down to 3.68 THz (at $F_2$ = 64 μJ/cm$^2$) as the value of $F_2$ is increased [Figs. 5(b) and 5(c)]. Here the FT spectra in Fig. 5(b) are obtained from the time-domain data in Fig. 5(a) just after the second pump-pulse to exclude any artifact originated from the periods of the transient electronic response induced by the pump-pulse pair.

In the context of Figs. 5(b) and 5(c), a vibrational frequency shift due to anharmonic effects would play a minor role, since the pump fluence is less than $\approx 10^{-2}$ of that needed to induce coherent anharmonic vibrations [26]. Furthermore, the origin of the frequency shift is not lattice heating, because the lattice heating should generally occur on several picosecond time scales [27]. Consequently, the frequency shift, from $\Omega_A$ to $\Omega_C$, observed by enhancing the vibrational amplitude is attributed to the ultrafast crystallization. Note that it is natural that the $A_1$ mode at $\approx$ 5.0 THz (SbTe$_3$ mode) is weakly preserved after the crystallization, because the GST-SL film holds the pyramidal SbTe$_3$ bonds in both phases if the mixing of sub-layers does not occur [18, 19, 21].

As the second procedure, the dependence of the phonon frequency on the time interval $\Delta t$ of the pump-pulse pair is an intriguing way to explore whether the ultrafast phase change in GST-SL film is stimulated by a coherent phenomenon. Here, the fluences of the first and the second pump-pulses were fixed at 76 and 64 μJ/cm$^2$, respectively, and $\Delta t$ was varied from 141 to 276 fs. For the shortest time interval [$\Delta t$ =141 fs in Fig. 5(d)], the amplitude of the coherent vibrations is suppressed by the second pump-pulse because of the destructive interference between the firstly and the secondly generated coherent vibrations by each pump-pulse [23, 25]. On the contrary, as $\Delta t$ increases up to 276 fs, the signal is enhanced after the irradiation of the second pump-pulse due to the constructive interference [23, 25]. The corresponding FT spectra [Fig. 5(e)] revealed a red-shift of the peak frequency, depending on the value of $\Delta t$; the frequency of the $A_1$ mode decreases toward that of the crystalline phase ($\approx$ 3.68 THz) as $\Delta t$ approaches 276 fs [Fig. 5(f)]. These data strongly indicate that the coherent (constructive) excitation of the local $A_1$ mode is crucial to induce the ultrafast phase change.

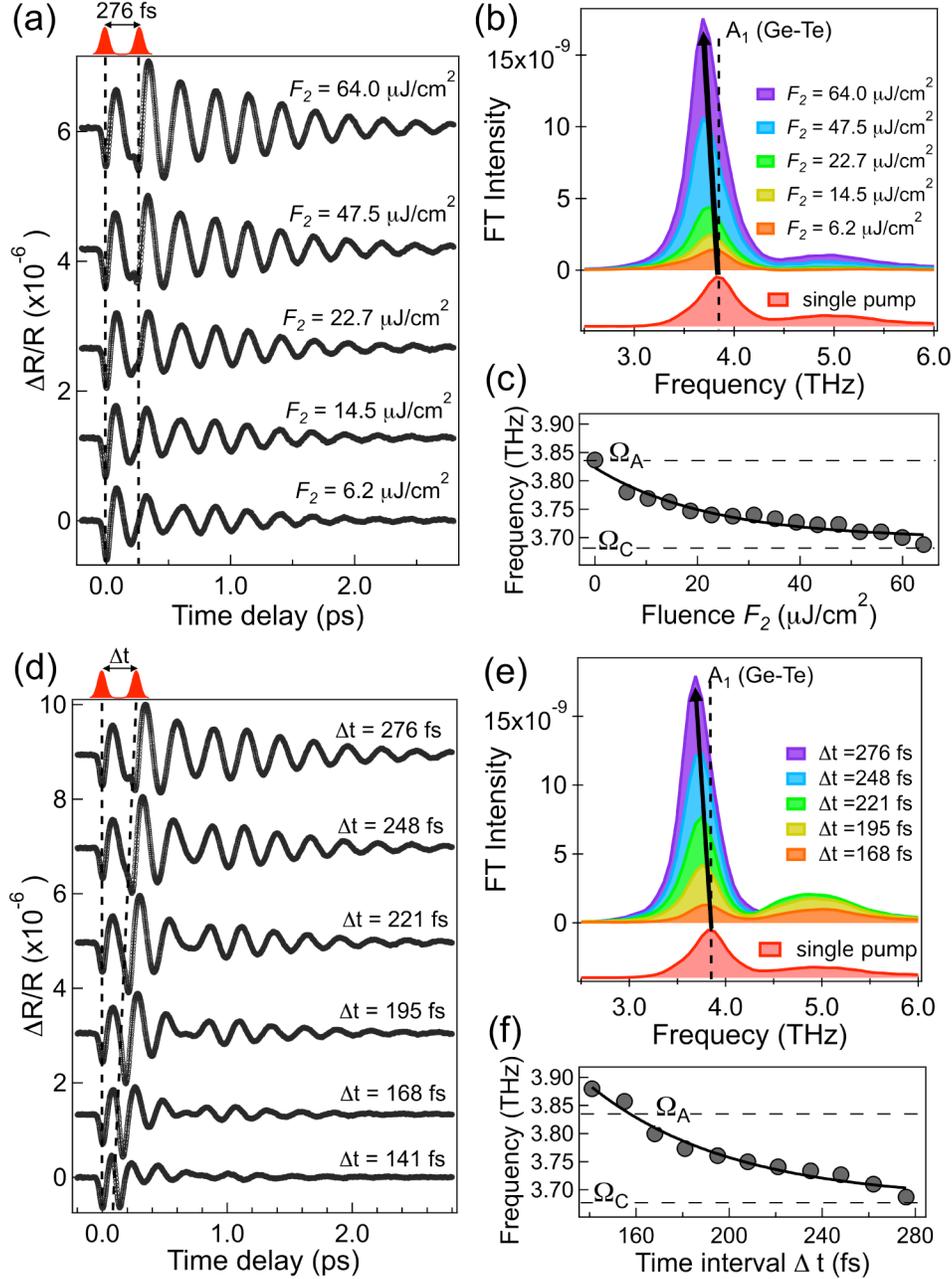

Fig. 5. (a) Coherent vibrations in the amorphous phase excited by the pump-pulse pair as the fluence of the second pump-pulse ($F_2$) is varied. The primary pump-pulse fluence was fixed at $F_1 = 76$ μJ/cm$^2$, while $F_2$ was tuned from 0 (without second pump) to 64 μJ/cm$^2$. (b) FT spectra from the data in (a). The arrow indicates the shift of the local $A_1$ mode (GeTe$_4$), which is more visible in the variation of the frequency as the function of $F_2$ in (c). The dotted lines in (c) correspond to the frequencies of the local $A_1$ mode (GeTe$_6$) in the crystalline phase (= 3.68 THz) and that (GeTe$_4$) in the amorphous phase (= 3.83 THz), respectively. The solid curve is the guide for the eyes. (d) Coherent vibrations excited by the pump-pulse pair as $\Delta t$ is varied. (e) FT spectra from the time-domain data in (d). The FT spectrum at $\Delta t = 141$ fs is omitted since the intensity is too low to display. The arrow indicates the peak-shift of the local $A_1$ mode (GeTe$_4$), being more visible in (f), which represents the frequency of the local $A_1$ mode as the function of the time interval $\Delta t$. The solid curve is the guide for the eyes.

## 6. Time dependent frequency of the coherent $A_1$ mode

However, the most intriguing aspect of the coherent lattice response under the irradiation of the pump-pulse pairs is a transient softening of the frequency, which occurs within 1000 fs (= 1 ps) after the irradiation of the second pump-pulse (Fig. 6). When the value of $F_2$ is set to be zero (single pump-pulse excitation), the transient frequency of the coherent local $A_1$ mode (GeTe$_4$) initially drops toward the minimum in the vicinity of 3.6 THz at ≈ 400 fs and finally returns to nearly the original value within 1000 fs. These ultrafast dynamics can be explained by the fact that the vibrational frequency in the excited electronic states is softened even if the pump fluence is quite low, and it returns to the original one after a certain time delay [28]. On the contrary, once the second pump-pulse is cooperated, the transient frequency of the $A_1$ mode decreases as well, but it is finally quenched at a much lower frequency than that in the case of the single pump-pulse experiment, depending on the value of $F_2$ (see the middle and the bottom traces in Fig. 6). Consequently, the time-dependent frequency-shift demonstrates that the phase change in the GST-SL film can be manipulated within a time scale of 1 picosecond by precisely controlling the parameters of the second pump-pulse.

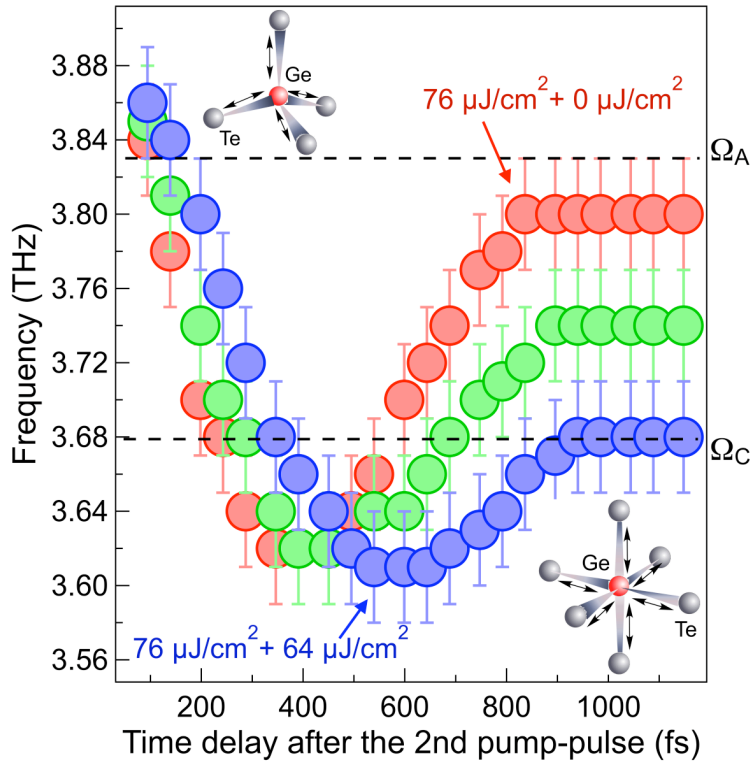

Fig. 6. Transient frequency of the coherent $A_1$ mode obtained by a wavelet analysis from the time-domain signal with single and pump-pulse pair excitations. The red circles represent the case of the single pump-pulse excitation ($F_2 = 0$ μJ/cm$^2$), while the green circles are for the pump-pulse pair excitation with $F_2 = 14.5$ μJ/cm$^2$, and the blue circles are for those with $F_2 = 64.0$ μJ/cm$^2$.

It is to be noted that although the observation of the time-dependent frequency (Fig. 6) originates from the transient structural dynamics in femtosecond time scale as described above, we have noticed that the GST-SL film permanently changed into crystalline after the femtosecond laser exposure over 1 hour. A single-shot to multi-shots pump-probe measurements will be useful to investigate how many shots of the laser pump-pulse pairs are required for the permanent phase change.

**7. Discussions**

We discuss why two pump-pulses are needed to realize the ultrafast phase change in GST-SL. Recently, Kolobov *et al.* reported that the rapid phase change between the amorphous and crystalline phases was characterized by a flip-flop transition of Ge atoms, that is a displacement of Ge atoms from tetrahedrally Te-coordinated site to octahedrally Te-coordinated site [8]. More recent molecular dynamics simulations have suggested in the amorphous phase of GST that the majority of the atoms have tetrahedral atomic arrangement ($GeTe_4$) with about 40–50 % occupations [10].

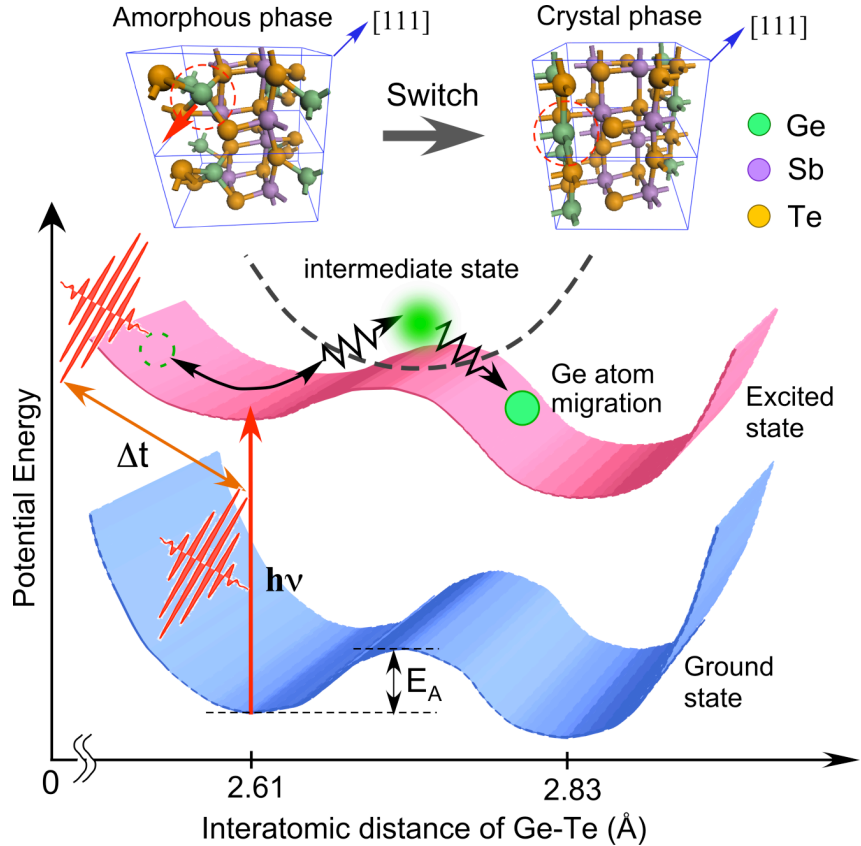

Fig. 7. Schematic of the potential energy surface during the pump-pulse pair excitation, together with the local structural change of GST-SL (Top). $E_A$ represents the activation energy when the phase change is promoted by thermal (incoherent) process. The position of Ge atoms can be rearranged by the coherent selective excitation of the local vibration, $GeTe_4$, accompanying with surrounding (Sb and Te) atomic rearrangement via an intermediate state (dashed curve). Top: the local structures in the two phases were calculated by the first principle simulation based on density functional theory (DFT). The dashed red circles highlight the local structures of $GeTe_4$ (in the amorphous phase) and $GeTe_6$ (in the crystalline phase). The red arrow represents the atomic displacement of the Ge atom along the [111] direction.

Hence, our results suggest that the pump-pulse pair selectively modulates the bonding of Ge-Te in the $GeTe_4$ units in the amorphous state, making Ge atoms migrate into octahedral position along the [111] direction, accompanying with surrounding (Sb and Te) atomic rearrangement (Fig. 7). This is possible because in the excited states one can reconfigure chemical bonding easier than that in the ground states [29]. Furthermore, using first-principle simulations within the local density approximation, availability of the phase transition driven by the displacement of Ge atoms along the rocksalt [111] direction has been pointed out for the system of $(GeTe)_2/(Sb_2Te_3)$ [18] (see also the top of Fig. 7), supporting our interpretations above.

Note that these findings in femtosecond time scale have unfortunately been masked for a long time when using picosecond laser pulses in GST alloy films [30].

## 8. Conclusions

We have experimentally shown that the ultrafast phase change from amorphous into crystalline phases in GST-SL films is manipulated using the femtosecond pump-pulse pairs, which is easily accessed by a fiber-type compact femtosecond laser oscillator as well. Future requirements for memory devices include larger capacities, significantly shorter access times, smaller physical dimensions and more practically lower power operations [31]. Our demonstration of nonthermal ultra-high-speed atomic rearrangements in the GST-SL films coupled with ultra-low fluence laser pulses may provide highly relevant for the next-generation of ultra-high-speed phase change memory (PCRAM). Furthermore, if an ultra-high-speed optical switch operating at the terahertz frequencies [32] in PCRAM devices is realized, this technique will revolutionize optical switching with the ultrabroad-band conversion efficiency quintessential for optical communications and at the same time allow realization of ultrafast data transfer rate with Tbit/s speed far beyond the current speed limit of ~ 160 Gbit/s.

## Acknowledgments

The authors acknowledge P. Fons and A. V. Kolobov for critically reading manuscript and stimulating discussions. The authors thank Y. Miyamoto for the assistance at the early stage of the experiments and D. Hayashi for the construction of the interferometer. This work was supported in part by KAKENHI-22340076 from MEXT, Japan and by the "Innovation Research Project on Nanoelectronics Materials and Structures - Research and development of superlattice chalcogenide phase-change memory based on new functional structures" from METI, Japan.